\documentclass[12pt,preprint]{aastex}
\renewcommand\>{{\rangle}}
\newcommand\bld[1]{\mbox{\boldmath $#1$}}
\newcommand\<{{\langle}}

\newcommand{\pdv}[2]{\frac{\partial#1}{\partial#2}}
\newcommand{\bnabla}{\bld{\nabla}}
\newcommand{\bB}{\bld{B}}
\newcommand{\bx}{\bld{x}}
\newcommand{\by}{\bld{y}}
\newcommand{\bz}{\bld{z}}
\renewcommand{\bv}{\bld {v}}

\newcommand{\bO}{\bld{\Omega}}
\newcommand{\uv}[1]{\hat{\bld{#1}}}
\newcommand{\vorb}{\bv_{orb}}
\newcommand{\athena}{{\tt ATHENA }}
\newcommand{\athenap}{{\tt ATHENA}}
\newcommand{\zeus}{{\tt ZEUS }}
\newcommand{\zeusp}{{\tt ZEUS}}

\newcommand{\be}{\begin{equation}}
\newcommand{\ee}{\end{equation}}
\newcommand{\bea}{\begin{eqnarray}}
\newcommand{\eea}{\end{eqnarray}}

\shortauthors{Guan et al.}
\shorttitle{}
\begin{document}

\title{Locality of MHD Turbulence in Isothermal Disks}

\author{Xiaoyue Guan and Charles F. Gammie\altaffilmark{1}}
\affil{Astronomy Department, University of Illinois,
1002 West Green St., Urbana, IL 61801, USA}

\author{Jacob B. Simon}
\affil{Astronomy Department, University of Virginia, Box 400325, Charlottesville, VA 22904, USA}

\and

\author{Bryan M. Johnson}
\affil{Lawrence Livermore National Laboratory, L-023, 7000 East Avenue, Livermore, CA 94550}

\altaffiltext{1}{Physics Department, University of Illinois}

\begin{abstract}

We numerically evolve turbulence driven by the magnetorotational
instability (MRI) in a 3D, unstratified shearing box and study its
structure using two-point correlation functions.  We confirm Fromang \&
Papaloizou's result that shearing box models with zero net magnetic flux
are not converged; the dimensionless shear stress $\alpha$ is
proportional to the grid scale.  We find that the two-point correlation
of $\bB$ shows that it is composed of narrow filaments that are swept
back by differential rotation into a trailing spiral.  The correlation
lengths along each of the correlation function principal axes decrease
monotonically with the grid scale.  For mean azimuthal field models,
which we argue are more relevant to astrophysical disks than the zero
net field models, we find that: $\alpha$ increases weakly with
increasing resolution at fixed box size; $\alpha$ increases slightly as
the box size is increased; $\alpha$ increases linearly with net field
strength, confirming earlier results; the two-point correlation function
of the magnetic field is resolved and converged, and is composed of
narrow filaments swept back by the shear; the major axis of the
two-point increases slightly as the box size is increased; these results
are code independent, based on a comparison of \athena and \zeus runs.
The velocity, density, and magnetic fields decorrelate over scales
larger than $\sim H$, as do the dynamical terms in the magnetic energy
evolution equations.  We conclude that MHD turbulence in disks is
localized, subject to the limitations imposed by the absence of vertical
stratification, the use of an isothermal equation of state, finite box
size, finite run time, and finite resolution.

\end{abstract}


\section{Introduction}

Astrophysical disks appear to redistribute angular momentum rapidly,
much more rapidly than one would expect based on estimates of the
molecular viscosity.  Classical thin accretion disk theories
\citep{ss,lbp} solved this problem by appealing to turbulence, and
modeled the effects of this turbulence as an ``anomalous viscosity.''
The idea that turbulence plays a key role was placed on firmer
foundations with the (re)discovery of the magnetorotational instability
(MRI; \citealt{bh91}) and subsequent numerical investigations (see
\citealt{bh98} for a review).  Winds or gravitational instability may
drive disk evolution in certain cases, but MRI-initiated MHD turbulence
appears capable of driving disk evolution in a wide variety of
astrophysical disks.

We still do not know, however, whether the effects of MHD turbulence on
disks are localized.  It is possible that structures develop that are
large compared to a scale height $H \equiv c_s/\Omega$, and that these
structures are associated with nonlocal energy and angular momentum
transport.  If so, disk evolution would not be well described by a
theory, such as the $\alpha$ model, in which the shear stress depends
only on the local surface density and temperature.

A related possibility, which we will not examine here, is that the
time-averaged turbulent stresses $\overline{W}_{r\phi}$ might satisfy
$\partial \overline{W}_{r\phi}/\partial \Sigma < 0$ ($\Sigma \equiv$
surface density; see \cite{pir78} for a discussion).  That is, the disk
might be ``viscously'' unstable.  This could cause the disk to break up
into rings or even---to use a term of art---``blobs.'' Such an outcome
would be awkward for the classic phenomenological steady disk and disk
evolution models, which have had some success in modeling cataclysmic
variable disks and black hole x-ray binary disks in a high, soft state
(e.g.  \citealt{belloni97, lasota01}).  

How can one probe the locality of MHD turbulence in disks?  We will use
the two-point correlation function of the magnetic field, velocity
field, and density as determined by numerical experiments.  Nonlocal
transport would likely be associated with features in the two-point
correlation function, as would viscous instabilities.  For example,
turbulence might excite waves (wakes) that carry energy and angular
momentum over many $H$ in radius before damping.  These wakes would
appear as extended features in the two-point correlation function.  

The two-point correlation function and the power spectrum contain the
same information since they are related by a Fourier transform.  But
they do not convey the same impression and they have different noise
properties.  For a one dimensional function sampled at $N$ points over
an interval $L$ half the sample points in the power spectrum lie between
the Nyquist frequency ($\pi N/L$) and half the Nyquist frequency,
while for the correlation function half the sample points lie between
a separation $L/4$ and $L/2$. The two point correlation function will
therefore convey a more accurate impression of large scale features than power spectra.

In this paper we study models with both zero net field and net azimuthal
field.  We ignore mean vertical field models because we remain persuaded
by the phenomenological argument of \cite{vanball} that vertical field
diffuses easily out of the disk when the turbulent magnetic Prandtl
number is $O(1)$ (although there are ways of evading this argument
\citep{uzspruit}).  Net azimuthal field models are, we think, most
relevant to astrophysical disks.  In disk galaxies---the only
differentially rotating disks where we can resolve field structure---the
azimuthal field dominates when averaged over areas more than a few $H^2$
in extent (e.g. Beck 2007).  Azimuthal field also dominates in global
disk simulations (e.g. \cite{hkvh05,mn07,bhk08}), and in local disk
simulations.  In local simulations in which the mean field is allowed to
evolve (e.g. \cite{bran95, ms00}) an azimuthal mean field develops.
Taken together these simulations and observations strongly suggest that
the azimuthal field averaged over regions $\sim H^2$ in area will never
be exactly zero.  

The paper is organized as follows.  In \S 2 we give a simple description
of the local model and summarize our numerical algorithm with orbital
advection.  We then study zero net flux models (as in Fromang \&
Papaloizou 2007; hereafter FP07) ; this serves as a code test and
introduces the correlation function analysis.  In \S 3  we explore the
properties of turbulence in models with a mean azimuthal field.  We
report on the saturation level and correlation lengths and we discuss
their dependence on the model parameters, such as resolution, box size,
and initial field strength. \S 4 contains a summary and guide to our
results.

\section{Model, Methods, and Tests}

Our starting point is the local model for disks.  It is obtained by
expanding the equations of motion around a circular-orbiting coordinate
origin at cylindrical coordinates $(r,\phi,z) = (r_o, \Omega_o t +
\phi_o, 0)$, assuming that the peculiar velocities are comparable to the
sound speed and that the sound speed is small compared to the orbital
speed.  The local Cartesian coordinates are obtained from cylindrical
coordinates via $(x,y,z) = (r - r_o, r_o [\phi - \Omega_o t - \phi_o],
z)$.  We assume throughout that the disk is isothermal ($p = c_s^2
\rho$, where $c_s$ is constant), and that the disk orbits in a Keplerian
($1/r$) potential.

In the local model the momentum equation of ideal MHD becomes
\be\label{BE2}
\pdv{\bv}{t} + \bv\cdot \bnabla \bv + c_s^2\frac{\bnabla\rho}{\rho} + \frac{\bnabla B^2}{8\pi \rho}
- \frac{(\bB\cdot \bnabla)\bB}{4\pi \rho} + 2 \bO \times \bv - 3\Omega^2 x \, \uv{x} = 0.
\ee
The final two terms in equation (\ref{BE2}) represent the Coriolis and
tidal forces in the local frame.  Notice that our model is unstratified,
which means that the vertical gravitational acceleration $-\Omega^2 z$
usually present in Keplerian disks is ignored.

Our model contains no explicit dissipation coefficients.  Recent models
with explicit scalar dissipation (FP07, \citealt{ll07}) have shown that
the outcome (saturated field strength) depends on the viscosity $\nu$
and resistivity $\eta$, and that \zeus has an effective magnetic Prandtl
number $Pr_M \equiv \nu/\eta \sim 4$.  

The orbital velocity in the local model is
\be
\vorb = -{3\over{2}}\Omega x \, \uv{y}.
\ee
This velocity, along with a constant density and zero magnetic field, is
a steady-state solution to equation (\ref{BE2}).  If the computational
domain extends to $|x| > (2/3) H = (2/3) c_s/\Omega$, then the orbital
speed is supersonic with respect to the grid.

The local model is studied numerically using the ``shearing box''
boundary conditions (e.g. \citealt{hgb95}).  These boundary conditions
isolate a rectangular region in the disk.  The azimuthal ($y$) boundary
conditions are periodic; the radial ($x$) boundary conditions are
``nearly periodic'', i.e. they connect the radial boundaries in a
time-dependent way that enforces the mean shear flow.  We use periodic
boundary conditions in the vertical direction; this is the simplest
possible version of the shearing box model.

\subsection{Numerical Methods}

Most of our models are evolved using \zeus \citep{sn92}.  \zeus is an
operator-split, finite difference scheme on a staggered mesh.  It uses
artificial viscosity (not an anomalous viscosity!) to capture shocks.
For the magnetic field evolution \zeus uses the Method of
Characteristics-Constrained Transport (MOC-CT) scheme, which is
designed to accurately evolve Alfv\'en waves (MOC) and also to preserve
the $\bnabla \cdot \bB = 0$ constraint to machine precision (CT).

We have modified \zeus to include ``orbital advection'' \citep{mass00,
gamm01, jg05} with a magnetic field \citep{jgg08}.  Advection by the
orbital component of the velocity $\bv_{orb}$ (which may be supersonic
with respect to the grid) is done using interpolation.  With this
modification the timestep condition $\Delta t < {\it C}  \Delta
x/(|\delta \bv| + c_{max})$ ($c_{max} \equiv $ maximum wave speed and
${\it C} \equiv$ Courant number) depends only on the perturbed velocity
$\delta \bv = \bv - \bv_{orb}$ rather than $\bv$.  So when $|\bv_{orb}|
\gtrsim c_{max}$ (for shearing box models with $v_A^2/c_s^2 \lesssim 1$,
when $L \gtrsim H$) the timestep can be larger with orbital advection,
and computational efficiency is improved.

Orbital advection also improves accuracy. \zeusp, like most Eulerian
schemes, has a truncation error that increases as the speed of the fluid
increases in the grid frame.  In the shearing box without orbital
advection the truncation error would then increase monotonically with
$|x|$.  Orbital advection reduces the amplitude of the truncation error
and also makes it more nearly uniform in $|x|$ \citep{jgg08}.  

Do our results depend on the algorithm used to integrate the MHD
equations?  To find out, we have also evolved a subset of models using
\athenap, a second-order accurate Godunov scheme that solves the
equations of ideal MHD in conservative form.  The algorithm couples the
dimensionally unsplit corner transport upwind (CTU) method of \cite{c90}
with the third-order in space piecewise parabolic method (PPM) of
\cite{cw84} and a constrained transport (CT) algorithm for preserving
the $\bnabla \cdot \bB = 0$ constraint.  Details of the algorithm and
test problems are described in \cite{sgths08}.  The specific application
of \athena to the shearing box (as used in this work) is described in
\cite{shb08}.\footnote{\cite{shb08} use $P = (\gamma - 1) u$ in contrast
to our $P = c_s^2 \rho$.}

\subsection{Models with Zero Net Flux }

We now consider a set of zero net field shearing box models to introduce
our correlation function analysis.  We use the same model parameters as
FP07 so that these models also serve, by comparison with FP07, as a
nonlinear code test.

The models in this section have size $(L_x, L_y, L_z) = (1, \pi, 1) H$.
The initial magnetic field is $B_z = B_{z0} \times \sin(2 \pi x/H)$,
where $B_{z0}$ satisfies $\beta \equiv 8\pi P_0/B_{z0}^2 = 400$.  Noise
is introduced in the initial velocity field to stimulate the growth of
the unstable modes.  The models are evolved to $t_f = 600 \Omega^{-1}$. 

\begin{figure}
\plotone{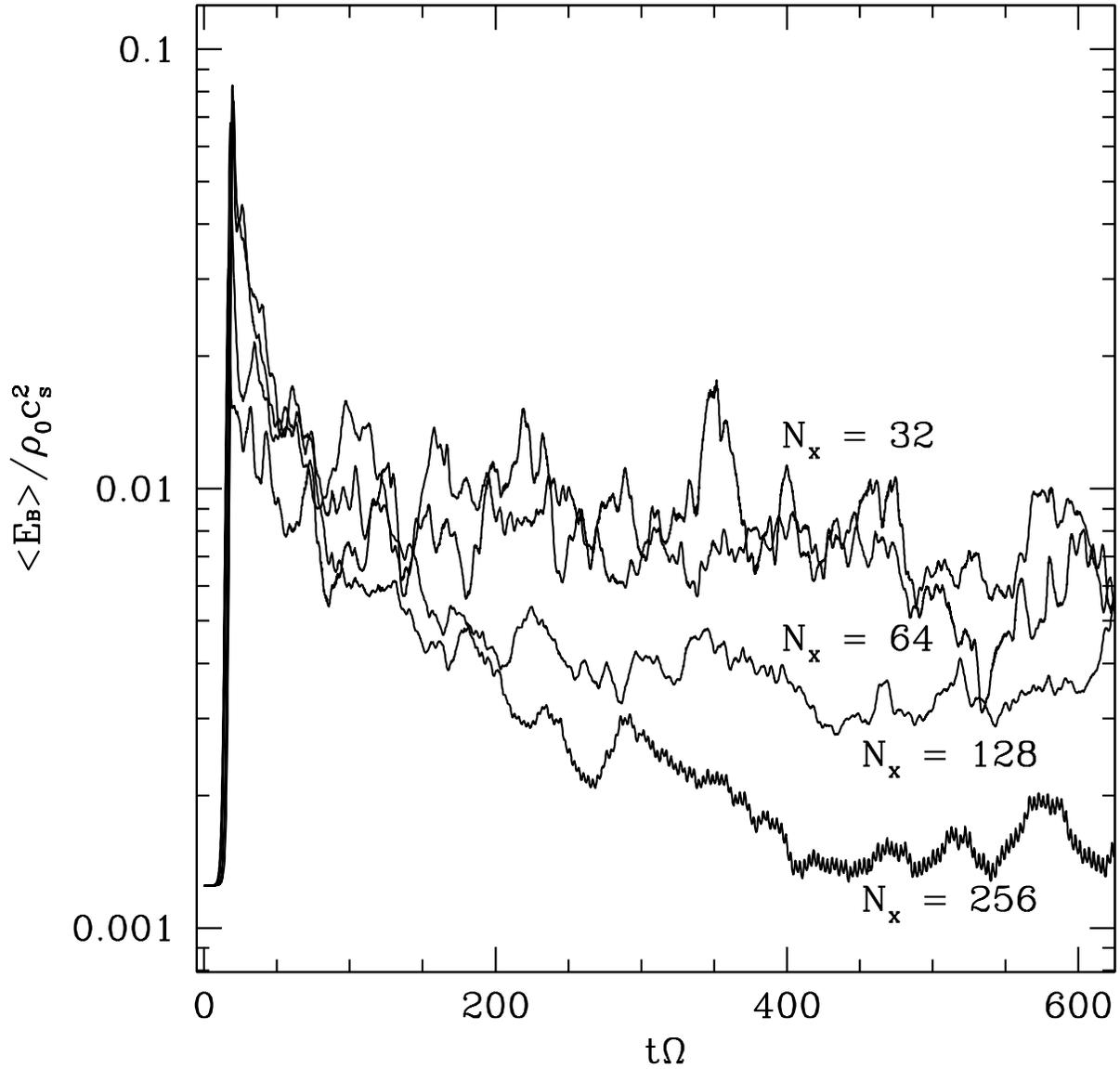}
\caption{
Evolution of magnetic energy in zero-net field runs. From top to bottom
radial resolution increases from $32/H$, $64/H$, $128/H$ to $256/H$. The
saturation level decreases in proportion to the grid scale.
}
\label{fig:std.eb.vs.t}
\end{figure}

We consider four resolutions: $(N_x, N_y, N_z) = N (32, 50, 32)$, where
$N = 1, 2, 4, 8$.  The last three models correspond to runs std32, std64
and std128 in FP07, respectively.  The evolution of $\<E_B\> \equiv
\<B^2/(8\pi \rho_o c_s^2)\>$ is shown in Figure \ref{fig:std.eb.vs.t}
($\<\> \equiv$ volume average).  The saturation $\<E_B\>$ decreases as
the resolution increases.

The dimensionless shear stress
\begin{equation}
\alpha \equiv {\<\rho v_x \delta v_y  - {B_x B_y\over{4\pi}} \>
	\over{\<\rho\> c_s^2}}.
\end{equation}
We measured the time average of $\alpha$, denoted $\overline{\alpha}$,
(from $t\Omega = 250$ to $t\Omega = 600$) for each run, and recorded the
results in Table \ref{std}\footnote{The combination of finite run time
and fluctuations in $\alpha(t)$ introduce noise into
$\overline{\alpha}$.  To estimate the noise amplitude we divided the
averaging interval into $2$ and compared the two averages.  In all the
runs with $N_x \ge 64$ case they differed from the mean by $\leq
10\%$.}.  These averages are nearly identical to those obtained by FP07.
This consistency enhances confidence in both sets of results. 

For $N_x \ge 64$ 
\begin{equation} 
\overline{\alpha} \simeq 0.0021 \left({{N_x}\over{128}}\right)^{-1}
\end{equation} 
is a good fit to the numerical results.  The magnetic field energy
density, $\alpha$, and the kinetic energy density are almost inversely
proportional to $N_x$, and so, like FP07, we conclude that the zero net
field models do not converge.

All shearing box models considered in this paper have $\overline{\alpha}
\simeq \overline{\<E_B\>}/2 = 1/(2\overline{\beta})$.  This implies a
characteristic orientation of the field, since (neglecting the Reynolds
stress $\rho v_x v_y \sim 0.25 \times [-B_x B_y]/[4\pi]$)
$\overline{\alpha} \approx - \overline{\<B_x B_y\>}/(4\pi \<\rho\>
c_s^2) \equiv \overline{\< B^2 \cos\theta_B \sin\theta_B \>}/(4\pi
\<\rho\> c_s^2)$, where $\theta_B$ is the angle between the field and
the $y$-axis. Then $\overline{\alpha} = \overline{\<E_B\>} 2
\cos\overline{\theta}_B \sin\overline{\theta}_B = \overline{\<E_B\>}/2$.
So $\overline{\theta}_B = \pi/12$ ($15^\circ$) is the characteristic
angle between the magnetic field and the $y$-axis. 

\begin{figure}
\plotone{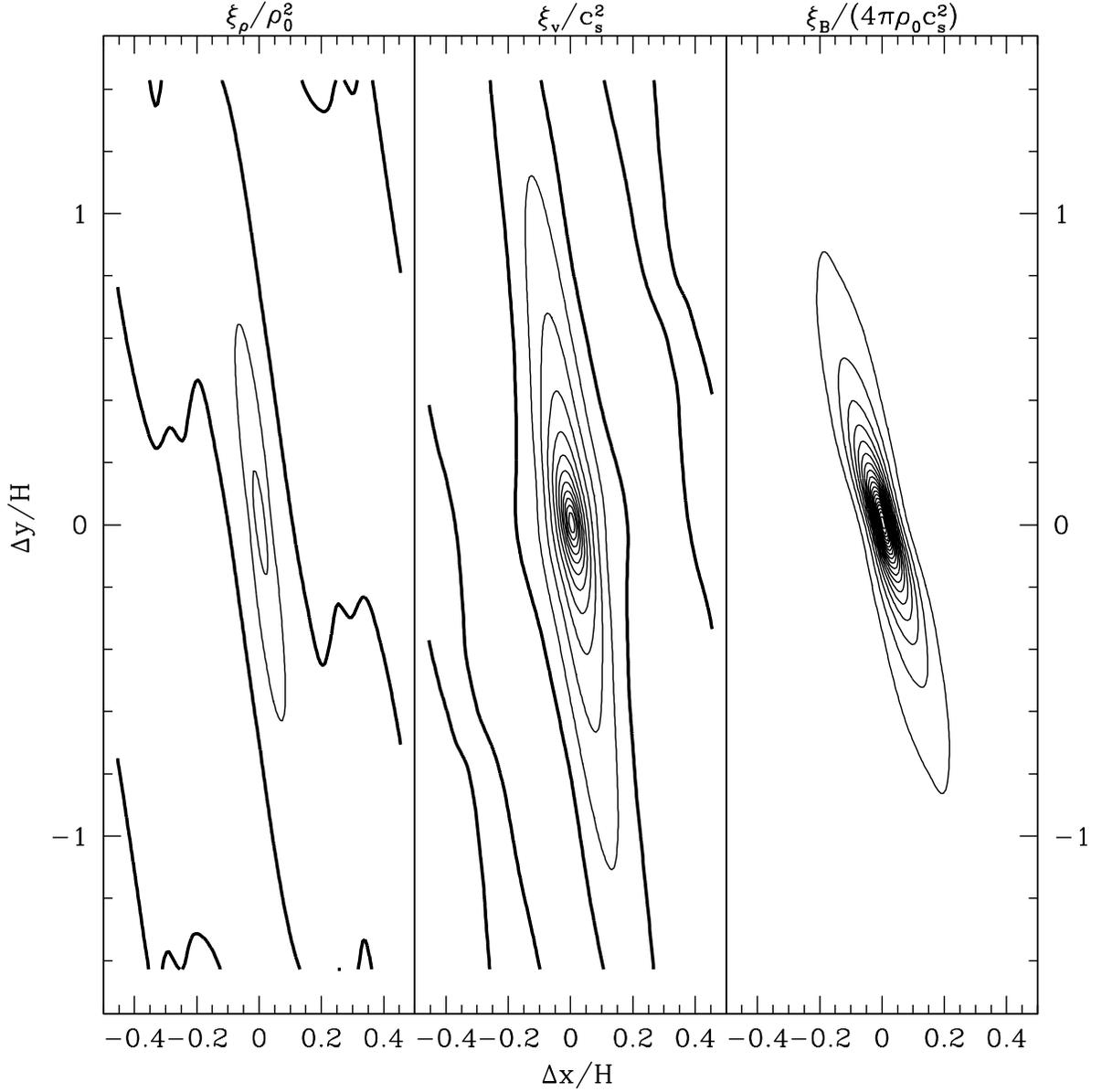}
\caption{
Two-point correlation function for density, velocity field and magnetic
field in $\Delta x-\Delta y$ plane in run $z128$. The contours are set
linearly from $0$ to $0.009$ for $20$ levels; the heavy line is the $0$ 
contour.
}
\label{fig:std.corr.3fig}
\end{figure}

Next we turn to the structure of the zero net field turbulence.
Consider the two-point correlation function for the density fluctuations
\begin{equation}
\xi_{\rho} \equiv \< \delta\rho(\bx) \delta \rho(\bx + \Delta \bx) \>
\end{equation}
($\delta\rho \equiv \rho - \<\rho\>$), for the trace of the velocity
fluctuation correlation tensor
\begin{equation}
\xi_{v} = \< \delta v_i(\bx) \delta v_i(\bx + \Delta \bx) \>
\end{equation}
($\delta v_i \equiv v_i - v_{i,orb} - \< v_i \>$), and there is an
implied summation over $i$) and for the trace of the magnetic field correlation tensor
\begin{equation}
\xi_B = \< \delta B_i(\bx) \delta B_i(\bx + \Delta \bx) \>
\end{equation}
($\delta B_i \equiv B_i - \< B_i \>$). All correlation functions are calculated for for fluctuating dynamical
variables with zero mean.  Figure \ref{fig:std.corr.3fig} shows slices
of correlation functions through $\xi$ at $z = 0$ for run $z128$.  The
cores of the correlation functions are ellipsoidal, with three principal
axes, and  concentrated at $|\Delta \bx| < H$.  The correlations are
localized.

We measure four features of the correlation functions: the angle
$\theta_{\rm tilt}$ between the correlation function major axis and the
$y$-axis, and the correlation lengths along the major, minor, and
$z$-axes ($\lambda_{\rm maj}, \lambda_{\rm min}, \lambda_{\rm z}$),
where the correlation length $\lambda_i$ is defined \footnote{Other
definitions of $\lambda_i$ are possible; e.g. the half width at half
maximum (HWHM) of $\xi$ can be used, with an exponential model for
$\xi$, to find $\lambda_i$.  For example, for $\xi_B$ in run z128, the
HWHM definition gives $(\lambda_{\rm maj}, \lambda_{\rm min},
\lambda_{\rm z}) = (0.031, 0.15, 0.023)H$ compared to $(0.026, 0.15,
0.022)H$ from our definition.  The differences are $< 20\%$.} by
\begin{equation}\label{CORLENDEF}
\lambda_i \equiv {1\over{\xi(0)}} \int _{0}^{\infty} \xi(\hat{\bx}_i l)
dl. 
\end{equation}

\begin{figure}
\plotone{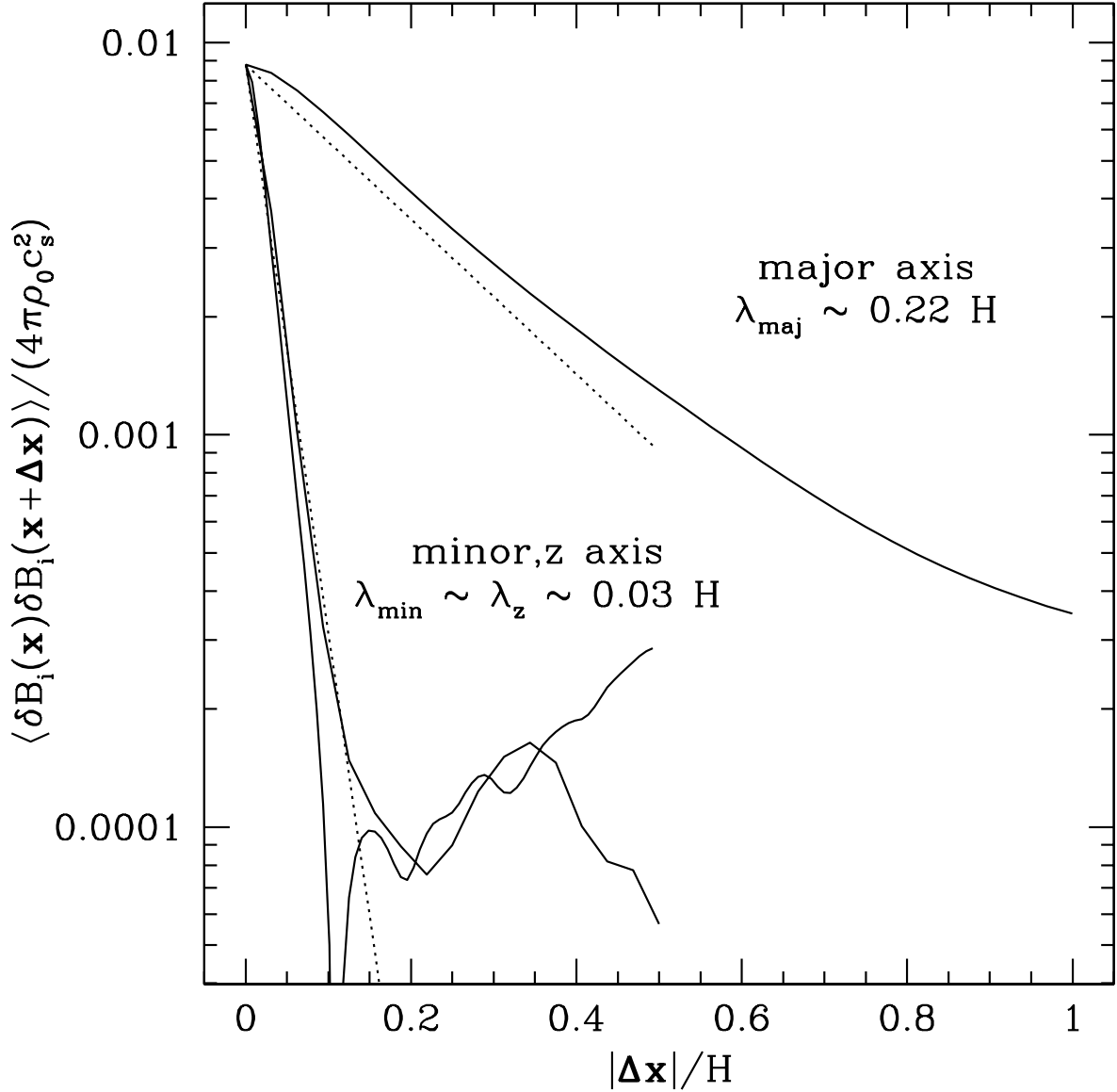}
\caption{
Magnetic field correlation function along the minor, major and vertical
principle axes in run $z128$. Solid lines: cut through the data; dotted
lines: a simple model with $\exp(-\lambda _{i})$, where $\lambda_{i}$ is
the measured correlation length along each principle axis.  The
correlation functions along the minor and $z-$axes are almost
identical.}
\label{fig:std.corr.2ray}
\end{figure}

$l$ is the distance from $\Delta \bx = 0$ along the principal axis
defined by the unit vector $\hat{\bx}_i$, and $\hat{\bx}_{\rm maj} =
\hat{\bx} \sin\theta_{\rm tilt} - \hat{\by} \cos\theta_{\rm tilt}$,
$\hat{\bx}_{\rm min} = \hat{\bx} \cos\theta_{\rm tilt} + \hat{\by}
\sin\theta_{\rm tilt}$, $\hat{\bx}_z = \hat{\bz}$.  \footnote{In a
periodic domain $\int d^3\Delta x_f \xi = 0$ if $\int d^3x f = 0$,
but this does not imply that the line integral in (\ref{CORLENDEF}) 
vanishes.} \footnote{The integral in
(\ref{CORLENDEF}) is evaluated by linearly interpolating $\xi$ in the
$\Delta x - \Delta y$ plane and
summing over the interpolated values (trapezoidal rule) along the
principal axis.  We evaluate the line integral until $\xi(l) = e^{-3}
\xi(0)$.  The result is insensitive to the upper limit on the
integral.} Figure \ref{fig:std.corr.2ray} shows $\xi_B$ along each of
the principal axes in run $z128$.  The dotted lines show $\xi(0)
\exp(-l/\lambda_{i})$.  The correlated regions in the magnetic field are
narrow ($\lambda_{\rm min} \simeq \lambda_{\rm z} \simeq \lambda_{\rm
maj}/6$) filaments with a trailing spiral orientation.  

What do the correlation functions mean?  For the magnetic field, there
is a characteristic orientation of the field $\overline{\theta}_B$
obtained through our measurement of the shear stress.  The major axis of
the correlation function is very nearly parallel to this, $\theta_{\rm
tilt} \simeq \overline{\theta}_B$.  It is reasonable to view the
magnetic field correlation function, then, as tracing out a
characteristic, filamentary structure in the magnetic field.

It is worth recalling briefly what we might expect for $\xi_v$ in
isotropic, homogeneous turbulence with an outer scale $L_0 = 2 \pi/k_0$
and velocity dispersion $\sigma_v$.  In the inertial range $v_k^2
\propto k^{-11/3}$, so a reasonable functional form for the power
spectrum is
\begin{equation}
v_k^2 = N \left(\sigma_v^2 \over{k_0^3}\right) 
{1\over{(1 + (k/k_0)^2)^{11/6}}},
\end{equation}
where $N$ is a nondimensional normalization constant.  The corresponding
autocorrelation function is 
\begin{equation}\label{ISOCORRFUNC}
\xi_v = \sigma_v^2 {\sqrt{3} \Gamma(\frac{2}{3})\over{2^{1/3}\pi}} 
(k_0 r)^{1/3} {\rm K}_{1/3}(k_0 r)
\end{equation}
where $K$ is the modified Bessel function and $r \equiv |\Delta
x|$.  For $k_0 r \ll 1$,
\begin{equation}
\xi_v \approx \sigma_v^2 (1 - 0.955 \, (k_0 r)^{2/3} + O(r^2)).
\end{equation}
This is the usual $r^{2/3}$ Kolmogorov dependence at small separation.
For $k_0 r \gg 1$,
\begin{equation}
\xi_v \approx \sigma_v^2 \,\, 0.743 \,\, (k_0 r)^{-1/6} \, e^{-k_0 r}.
\end{equation}
which yields the expected decorrelation for $k_0 r \gg 1$.  The 
correlation length defined in (\ref{CORLENDEF}) is $0.838/k_0$.
Notice that the power spectrum does not have zero power as $k
\rightarrow 0$; rather it asymptotes to $N \sigma_v^2 k_0^{-3}$.

For MHD turbulence in disks, however, the correlation function is not
isotropic, and its structure is not anticipated by any predictive
theory.  In the absence of such a theory it may be useful to have a
convenient analytic representation of the numerical results.  This can
be obtained by stretching equation (\ref{ISOCORRFUNC}) along each of the
principal axes, that is, by replacing $k_0 r$ by $u \equiv ( (\Delta \bx
\cdot \hat{\bx}_{\rm min}/\lambda_{\rm min})^2 + (\Delta \bx \cdot
\hat{\bx}_{\rm maj}/\lambda_{\rm maj})^2 + (\Delta \bx \cdot
\hat{\bx}_{\rm z}/\lambda_{\rm z})^2)^{1/2}$.

Do the correlation lengths converge?  We find that the correlation
lengths are resolution dependent, with
\begin{equation}   
(\lambda_{\rm min}, \lambda_{\rm maj}, \lambda_{z} ) 
	\simeq (0.04, 0.24, 0.03) \, H \, 
	\left({{N_x}\over{128}}\right)^{-2/3};
\end{equation}
$\lambda_{\rm z}$ and $\lambda_{\rm min}$ are at most $6$ zones.  The
scaling of $\lambda_{\rm min}$ with zone size is clearly not linear, but
the $2/3$ power law scaling is just a fit to the data and should not be
taken too seriously.  The non-linear scaling does hint at the
possibility that, as resolution is increased and $\lambda_{\rm min}$ and
$\lambda_{\rm z}$ are better resolved, there could be a transition in
the outcome.

The major axis for $\xi_B$ lies $\sim 15^\circ$ from the $y-$axis.  For
the density and peculiar velocity field the tilt angle is $\sim
7^\circ$.  The latter tilt is consistent with the measured Reynolds
stress: $\<\rho v_x v_y\>/\<\rho v^2\> \sim 1/8$, so the average
perturbed velocity is tilted at $\sim 7^\circ$ to the $y-$axis.

Finally, notice that there are low-amplitude features in $\xi_v$ and
$\xi_\rho$ at scales of a few correlation lengths (see particularly in
Figure 4a).  These features may be due to the excitation of rotationally
modified sound waves by MHD turbulence.  To test this hypothesis notice
that, for tightly wrapped ($k_y \ll k_x$), linear sound waves
$\delta\rho_k^2/\rho_0^2 \simeq \delta v_k^2/c_s^2$.  A field composed
of these waves would then have $\xi_\rho/\rho_0^2 = \xi_v/c_s^2$.
Taking the differential correlation function $\xi_v/c_s^2
-\xi_\rho/\rho_0^2$ should therefore remove those pieces of $\xi_v$ that
are due to sound waves.  Figure 4b shows $\xi_v/c_s^2 -
\xi_{\rho}/\rho_0^2$ at $\Delta y = 0$.  Evidently much of the large-scale power is removed.  Figure \ref{fig:std.soundwave} shows another slice
through $\xi_v/c_s^2 - \xi_{\rho}/\rho_0^2$ at $\Delta z = 0$.  Again
the large-scale power is removed.  This is consistent with the
hypothesis that the largest-scale features in the correlation functions
are acoustic waves.  

\section{Models with a Net Azimuthal Field }

In this section we study models with $\<B_y\> \ne 0$. These models
correspond more closely to what is observed in shearing box models with
boundary conditions that allow the net field to evolve, what is seen in
global MHD models of disks, and what is observed in galactic disks than
the $\<\bB\> = 0$ models.

\subsection{Convergence}

To test convergence we use the same size models as the zero net field
runs, $(L_x, L_y, L_z) = (1, \pi, 1) H$.  $\<B_{y}\>$ is set so that
$\beta = 400$; in the initial conditions all other magnetic field
components vanish. The models are evolved to $t = 250 \Omega^{-1}$.  We
use five different resolutions: $(N_x, N_y, N_z) = N (32, 50, 32)$,
where $N = 1, 2, 4, 6, 8$.  In each run we average over the second half
of the evolution to measure $\overline{\<E_B\>}$ and
$\overline{\alpha}$.  We also measure the correlation lengths from
$\xi_\rho, \xi_v$, and $\xi_B$ using an average of the correlation
function calculated from $8$ data dumps in the second half of the run.
Parameters and results for these runs are listed in Table \ref{mb}.

Figure \ref{fig:mb.eb.vs.t} shows $\<E_B\>(t)$ for various resolutions.
In our two highest resolution runs, $\overline{\<E_B\>} = 3.7\times
10^{-2} \rho_{0} c_s^2 $ for run $y192a$ and $ 4.1\times 10^{-2}
\rho_{0} c_s^2$ for run $y256a$, respectively.  A consistent fit is
$\overline{\<E_B\>} \simeq 0.03 (N_x/128)^{1/3}$ for $32 < N_x < 256$.
The saturation energy increases with resolution.  It is unlikely that
this trend continues indefinitely.  As we will see below, the magnetic
field correlation lengths are unresolved at $N_x = 32$ but resolved at
$N_x = 256$.  This suggests that the increase in $\overline{\<E_B\>}$ is
caused by resolution of magnetic structures near the correlation length.
If there is little energy in structures much smaller than the
correlation length, then $\overline{\<E_B\>}$ should saturate at
somewhat higher resolution.

To check the algorithm-dependence of the results we ran an identical
set of models using \athenap.  The results are listed in Table \ref{mb}.
Both sets of models show a weak upward trend in $\overline{\<E_B\>}$
with resolution.  If one corrects for the approximately $2 \times$
higher effective resolution of \athena then the \athena and \zeus
results are quantitatively consistent with each other.

The correlation lengths for the zero net field runs do not converge.
What about the net azimuthal field models?  The magnetic field
correlation lengths are listed in Table 2 (other correlation lengths are
omitted for brevity, but they behave similarly).  For $N \ge 4$
both \zeus and \athena find
\begin{equation}
(\lambda_{\rm min}, \lambda_{\rm maj}, \lambda_{z} ) 
	\simeq (0.05, 0.32, 0.05) \, H.
\end{equation} 
For the $N \ge 4$ models $\lambda_{\rm min} \sim \lambda_{\rm z} >
8\Delta x$.  In the highest resolution ($N = 8$) \athena model,
$\lambda_{\rm min}/\Delta x \simeq 14$.  The correlation lengths are both
converged and resolved.

If MHD turbulence in disks has a forward energy cascade, as do three
dimensional hydrodynamic turbulence and the Goldreich-Sridhar model for
strong MHD turbulence in a homogeneous medium, then it is natural to
identify the correlation lengths with the outer, or energy injection,
scale.  Since $\lambda_{min} \simeq 15$ grid zones at our highest
resolution there is no resolved inertial range.  We anticipate that
future, higher-resolution numerical experiments with a mean azimuthal
field will show the development of an inertial range.

\subsection{Magnetic Energy Evolution}

At what scale is magnetic energy generated in MRI driven turbulence?  To
investigate this, we have studied the correlation function for each term
driving the evolution of the volume averaged magnetic energy:
\begin{equation}
\label{eb_eqn}
\dot{E_B} = 
- \langle \bnabla \cdot (\frac{1}{2}B^2{\bv})\rangle 
- \langle \frac{1}{2} B^2 \bnabla \cdot {\bv}\rangle 
+ \langle{\bB} \cdot ({\bB} \cdot \bnabla {\bv})\rangle - D
\end{equation}
where $D$ is the volume-averaged numerical dissipation rate.  The terms
on the right hand side can be interpreted as describing the effects of
advection, compression and expansion, field line stretching, and
numerical dissipation.  On average the first term is small (it should
vanish exactly for shearing box boundary conditions, but roundoff and
truncation error make it nonzero), and the third term dominates the
second by a factor of 20 in run y128b.  In a time and volume averaged
sense the right hand side must be zero, so numerical dissipation must
approximately balance energy injection by field line stretching.

Previous studies (FP07; \citealt{shb08}) analyzed a version of this
equation in the Fourier domain to study turbulent energy flow as a
function of length scale in the shearing box simulations.  Both studies
found that the magnetic energy is generated on all scales by the
background shear.  Here we perform a complementary analysis in the
spatial domain.  \footnote{ \cite{shb08} study the $k$ dependence of
analogous terms on the right hand side of an equation for
$\dot{|B_k|^2}$ (their eq. [19]), which scale like $B^2$.  We directly
autocorrelate the terms on the right hand side of (\ref{eb_eqn}), and
this scales like $B^4$.} 

We have computed the autocorrelation function for the field line
stretching term from run $y128b$ (without subtracting the mean).  Figure
~\ref{fig:me.xy.no_shear} shows the autocorrelation function at $\Delta
z = 0$.  From this figure we conclude that: (1) the scale and shape of
the correlation function is similar to that of the fundamental variables
($\bB$, $\bv$; recall that in y128b $(\lambda_{B, {\rm min}},
\lambda_{B, {\rm maj}}, \lambda_{B, {\rm z}}) = (0.058, 0.33, 0.051)H$); (2) energy is injected at
scales comparable to the correlation length of the fundamental
variables; (3) a superposition of magnetic structures similar to $\xi_B$
that are distributed with uniform probability in space would have a
power spectrum that is flat (white noise) at low $k$.  It is plausible
that terms in the Fourier-transformed magnetic energy equation would
also be flat at low $k$.  This would not imply that energy is injected
by dynamically meaningful structures at large scales; it would simply be
the consequence of an uncorrelated superposition of small, localized
features in the turbulence.

\subsection{Box Size}

Does the saturation energy or correlation length depend on the size of
the computational domain (box size)?  To investigate, we fix the
physical resolution at $64$ zones$/H$ and vary the model size: $(L_x,
L_y, L_z) = (1, \pi, 1) H$, $(2, \pi, 1) H$, $(1, 2\pi, 1) H$, $(2,
2\pi, 1) H$, ${\rm and} (4, 4\pi, 1) H$.  The model parameters and
outcomes are listed in Table \ref{size}.  

Evidently there is a weak dependence of $\overline{\<E_B\>}$ on box
size; it increases from $0.024\rho_{0}c_s^2$ for the smallest run to
$0.038\rho_{0}c_s^2$ for the largest run.  The magnetic field
correlation lengths also increase with box size, with $\lambda_{\rm maj}
= 0.36H$ for the smallest box (so $L_y/(2 \lambda_{\rm maj}) = 4.4$) to
$\lambda_{\rm maj} = 0.57H$ for the largest box (so $L_y/(2 \lambda_{\rm
maj}) = 11$).  This upward trend in correlation length is probably real,
but it is sufficiently small that it is difficult to separate from noise
in the correlation length measurements.

The correlation functions $\xi_\rho$ and $\xi_v$, unlike $\xi_B$, have
low amplitude tails extending out to the box size.  This can be seen in
Figure~\ref{fig:size.soundwave}, which shows $\xi_\rho$ and $\xi_B$ for
run $y64.x4y4$.  The tails are likely due to sound waves, and their
absence in the differential correlation function $\xi_v/c_s^2-
\xi_\rho/\rho_0^2$, shown in the middle panel of
Figure~\ref{fig:size.soundwave}, is consistent with this.

\subsection{Field Strength}

We now compare two models that differ only in their initial field
strength.  Both have a size  $L_x, L_y, L_z = (1, 2\pi, 1)H$ with
resolution $N_x, N_y, N_z = 64, 200, 64$.  One model has the same
initial field strength as our other models, $\beta_{0} = 400$, while the
other one starts with a stronger field, $\beta_{0} = 100$.

We found that the saturated magnetic energy for the $\beta_{0} = 400$
run is $\overline{\<E_{B}\>} = 0.035\rho_{0}c_s^{2}$. For the $\beta_{0}
= 100$ run $\overline{\<E_{B}\>} = 0.079\rho_{0}c_s^{2}$, slightly more
than twice the saturation level of the higher $\beta_{0}$ run.
Resolution may be playing a role here: the $\beta_0 = 100$ run has twice
the resolution per most unstable MRI wavelength, and we know the
saturation level depends on resolution when the field strength is
constant.  

Our results are consistent with the linear relation between
$\overline{\<E_B\>}$ and initial field strength for $\<B_y\> \ne 0$
models reported in Hawley, Gammie, \& Balbus (1995; hereafter HGB95)
(although it may also be consistent with a wide range of exponents for this relation).   Our results are
inconsistent with HGB95's claim that $\overline{\<E_B\>} \propto L_y$,
at least if the box size is $\gg \lambda_{\rm maj}$ (the good agreement
found for HGB95's predictor may be a coincidence).

At the level we can determine from two data points, our results are
consistent with $\overline{\<E_B\>} \propto \rho_0 c_s V_{A,y0}$ where
$V_{A,y0}$ is the initial azimuthal Alfven speed (i.e. here scale height
$c_s/\Omega$ replaces $L_y$ in HGB95; there are no other length scales
in the problem).  This is interesting: it implies that $\alpha$ depends
on the gas pressure, a result first reported by \cite{sano04} and thus compressibility
plays a role in saturation of the MRI!  

Our results suggest that $\overline{\<E_B\>}$ should scale differently
in compressible and incompressible models.  In the incompressible models
the only lengthscale available is the size of the box so in
incompressible models we must have $\overline{\<E_B\>} \sim \rho_0
(L\Omega)^a V_{A,y0}^b$, where $L$ is some combination of $L_x, L_y$,
and $L_z$, and the exponents $a$ and $b$ are not determined.  

The correlation lengths for the magnetic field in the two runs are
$(\lambda_{\rm min}, \lambda_{\rm maj}, \lambda_{z} ) \simeq (0.08,
0.45, 0.08)H$ for the $\beta_{0} = 400$ run and $(\lambda_{\rm min},
\lambda_{\rm maj}, \lambda_{z} ) \simeq (0.11, 0.58, 0.10)H $ for the
$\beta_{0} = 100$ run. To sum up, the correlation length increases
weakly as the initial field strength and the box size increases.  It is
not consistent with the scaling $\sim B_{y,0}/(\rho_0 \Omega)$ one would
expect if the correlation length scaled with the most unstable
wavelength of the background field, and it is not consistent with the
scaling $\sim \<B_y^2\>^{1/2}/(\rho_0\Omega) \sim B_{y,0}^{1/2}$ one
would expect if the correlation length is related to a characteristic
MRI length scale for the (larger) fluctuating field.

\section{Summary}

We have investigated the locality of MHD turbulence in an unstratified,
Keplerian shearing box model using the two-point autocorrelation function.  

We first considered models with zero net vertical field and
the same parameters as FP07.  Our slightly different orbital
advection algorithm reproduces earlier results on the relation between
the saturation level and resolution: zero net field models do not
converge.  

Consistent with this, we also find that as resolution increases the
correlation lengths for the velocity, density, and magnetic field
decrease.  A fit to the results yields the following scaling for the
magnetic field correlation lengths,
\begin{equation}   
(\lambda_{\rm min}, \lambda_{\rm maj},
\lambda_{z} ) \simeq (0.04, 0.24, 0.03)
	\left({{N_x}\over{128}}\right)^{-2/3}H,
\end{equation} 
i.e. the correlation length decreases as the resolution increases.

We then studied a set of models with net toroidal field and initial
$\beta = 400$.  These models are not completely converged in the sense
that they show a trend of increasing $\overline{\alpha}$ with
resolution.  They are converged in that the correlation lengths are well
resolved and constant near the highest resolution:
\begin{equation}
(\lambda_{\rm min}, \lambda_{\rm maj}, \lambda_{z} ) \simeq (0.05, 0.32,
0.05)H.
\end{equation}
But because $\lambda_{z}$ and $\lambda_{min}$ are only just resolved
(they are each at most 14 grid cells), we do not see an inertial range.
We expect that future higher-resolution models will show the development
of an inertial range.

We further examined the correlation function for the dominant (nonnumerical) field
line stretching term in the magnetic energy evolution
equation.  The correlation lengths are small compared to $H$, consistent with
the correlation lengths of the dynamical variables.  Evidently energy is
injected at scales comparable to or smaller than the correlation length.  

We also explored the influence of the box size on the outcome in the net
toroidal field models. We found a weak dependence on the box size but
only for $L_x \sim H$. This suggests that in shearing box simulations
the size of the box should be chosen to be at least a few scale heights
so that the correlation lengths are not ``squeezed'' by the boundary
conditions.  We also varied the initial field strength, and consistent
with earlier reports found that the saturation level
($\overline{\alpha}$ or $\overline{\<E_B\>}$) scales linearly with the
initial field strength.  Correlation lengths also increase as the field
strength increases, but not linearly.

So is disk turbulence really localized?  Our answer is mixed.  On the
one hand, our net toroidal field models have almost all the correlation
amplitudes contained within a region a few scale heights on a side and
in this sense the turbulence is indeed local.  On the other hand, we do
see signs of radiation of compressive waves by the turbulence in the
two-point correlation function.  These signs are most impressively
visible in Figure 4a, which shows vertically extended tails on the
density autocorrelation function in the $\Delta y = 0$ plane, and in
Figure 8, which shows azimuthally extended tails on the density
autocorrelation function in the $\Delta z = 0$ plane.  These tails are
matched by similar tails on the velocity autocorrelation function,
consistent with our hypothesis that they are due to the excitation of
rotationally modified sound waves.  

The influence of compressive waves excited by MHD turbulence on disk
evolution cannot, in the end, be assessed with the experiments and
analysis in this paper.  The key measurement needed is the radial
damping length of compressive waves due to absorption and scattering of
waves by turbulent eddies.  This would be most easily measured in a
separate experiment that studies the response of MHD turbulence to an
imposed sound wave.  Even this measurement would be incomplete because
it neglects additional damping related to stratification
\citep{linetal90,lo98}, but it would provide an upper limit on the
damping length.

We are studying the locality of MHD turbulence in a highly idealized
situation in which stratification and other aspects of the larger disk
--- such as the process that generates the imposed azimuthal magnetic
field, perhaps a global dynamo --- are absent.  Our unstratified model
is insensitive to some effects that could lead to the development of
global ($\lambda \sim R$) or mesoscale ($R \gg \lambda \gg H$)
structures.  Convection and rotation might reasonably be expected to
lead to dynamo activity manifesting itself as large scale structures in
the magnetic field.  Disk atmospheres might also develop large-scale,
coronal structures that delocalize disk evolution by transmitting
angular momentum and energy \citep{uzgoo}. 

Finite integration time and limited accuracy is also a concern.  It is
possible that large scale structures emerge only over hundreds of
rotation periods.  If the disk is subject to a ``viscous instability''
of the sort mentioned in the introduction then the timescale for growth
of a feature on scale $\lambda$ would be $(\alpha \Omega)^{-1}
(\lambda/H)^2$; this timescale could be hundreds of rotation periods for
the modest $\alpha$s seen in our models if the instability is damped for
$\lambda <$ few $\times H$.  Accurate, long duration, and expensive
integrations will be required in future searches for viscous
instability.

\acknowledgements

This work was supported by the National Science Foundation under grants
AST 00-93091, PHY 02-05155, and AST 07-09246, and a Sony Faculty
Fellowship, a University Scholar appointment, and a Richard and Margaret
Romano Professorial Scholarship to CFG.  Portions of this work were
completed while CFG was a Member at the Institute for Advanced Study
(2006-2007).  The authors are grateful to Shane Davis, Peter Goldreich,
Stu Shapiro, Fred Lamb, Friedrich Meyer, and John Hawley for
discussions.

\newpage

\begin{deluxetable}{lcccccccccccccc}
\setlength{\tabcolsep}{0.07in}
\tabletypesize{\scriptsize}
\tablecolumns{15}
\tablecaption{Shearing Box Runs with a Zero Net Vertical Field 
  \label{std}}
\tablewidth{0pt}
\tablehead{
\colhead{Model} & \colhead{resolution} &
\colhead{$\overline{\alpha}$} & \colhead{$\lambda_{B,{\rm min}}$} &
\colhead{$\lambda_{B,{\rm maj}}$}
& \colhead{$\lambda_{B,z}$} & \colhead{$\theta _{B,{\rm tilt}}$} &
\colhead{$\lambda_{v, {\rm min}}$}  & \colhead{$\lambda_{v,{\rm maj}}$}
& \colhead{$\lambda_{v,z}$} & \colhead{$\theta _{v,{\rm tilt}}$} &
\colhead{$\lambda_{\rho,{\rm min}}$} & \colhead{$\lambda_{\rho,{\rm maj}}$}
& \colhead{$\lambda_{\rho,z}$} & \colhead{$\theta _{\rho,{\rm tilt}}$}\\
\colhead{} & \colhead{} & \colhead{} & \colhead{} &
\colhead{} & \colhead{} & \colhead{} & \colhead{} & \colhead{} & \colhead{} & \colhead{} &
\colhead{} & \colhead{} & \colhead{} & \colhead{} }
\startdata
z32  & $32\times 50\times 32$     & $3.8\times 10^{-3}$ & 0.090 & 0.62
& 0.080  & 11   & 0.078 & 0.57 & 0.13 & 5.5 & 0.076 & 0.82 & 0.39 & 8.6
\\
z64  & $64\times 100 \times 64$   & $4.2\times 10^{-3}$ &  0.059 & 0.38 & 0.050  & 13
& 0.074 & 0.45 &0.18 & 7.1 & 0.056 & 0.60 & 0.33 & 6.6 \\
z128 & $128\times 200 \times 128$ & $2.1\times 10^{-3}$ & 0.037 & 0.22
& 0.032  & 14 & 0.053 & 0.32 & 0.11 & 6.7 & 0.043 & 0.62 & 0.33 & 6.4
\\
z256 & $256\times 400 \times 256$ & $1.1\times 10^{-3}$ & 0.024 & 0.17
& 0.024 & 14 & 0.035 & 0.24 & 0.10 & 5.9 & 0.019 & 0.34 & 0.18 & 5.8
\\

\enddata

\end{deluxetable}

\begin{deluxetable}{lccccccccc}
\setlength{\tabcolsep}{0.07in}
\tabletypesize{\scriptsize}
\tablecolumns{10}
\tablecaption{Shearing Box Runs with a Net Azimuthal Field 
  \label{mb}}
\tablewidth{0pt}
\tablehead{
\colhead{Model} & \colhead{algorithm} & \colhead{resolution} &
\colhead{$\overline{\alpha}$} &\colhead{$\overline{\< E_B \>}/\rho_{0}c_s^2$} &\colhead{$\lambda_{B,{\rm min}}$} &
\colhead{$\lambda_{B,{\rm maj}}$}
& \colhead{$\lambda_{B,z}$} & \colhead{$\theta _{B,{\rm tilt}}$} &
\colhead{$\lambda_{B,{\rm min}}/\Delta x$} \\
\colhead{} & \colhead{} & \colhead{} & \colhead{} &
\colhead{} & \colhead{} & \colhead{} & \colhead{} }
\startdata
y32a  & \zeus & $32\times 50\times 32$     & 0.0094 &
0.019 & 0.10 & 0.40 & 0.084  & 12  & 3.2 \\
y64a  & \zeus & $64\times 100 \times 64$   & 0.014 &
0.024 & 0.066 & 0.36 & 0.064 & 15 & 4.2 \\
y128a & \zeus & $128\times 200 \times 128$ & 0.015 & 0.028 & 0.053 & 0.28 & 0.049  & 15 & 6.8 \\
y192a & \zeus & $192\times 300 \times 192$ & 0.020 & 0.037 & 0.055  & 0.30   & 0.049  &  15 & 11 \\
y256a & \zeus & $256\times 400 \times 256$ & 0.021 & 0.041 & 0.049
& 0.27 & 0.045 & 15 & 13 \\
y32b  & \athena & $32\times 50\times 32$   & 0.018   &  0.032 & 0.11
& 0.63 & 0.09 & 15 & 3.3 \\
y64b  & \athena & $64\times 100 \times 64$  & 0.015   &  0.027 & 0.070
& 0.38 & 0.060 & 16 & 4.5\\
y128b & \athena  & $128\times 200 \times 128$ & 0.018 &  0.035 & 0.058
& 0.33 & 0.051 & 16 & 7.4\\
y192b & \athena &  $192\times 300 \times 192$ & 0.025 & 0.050 & 0.052
& 0.30 & 0.047 & 17 & 10\\
y256b & \athena & $256\times 400 \times 256$  & 0.027 & 0.055 & 0.053
& 0.32 & 0.049 & 16 & 14\\
 \enddata
\end{deluxetable}

\begin{deluxetable}{lccc}
\setlength{\tabcolsep}{0.07in}
\tabletypesize{\scriptsize}
\tablecolumns{4}
\tablecaption{Shearing Box Runs with a Net Azimuthal Field: Effect of
  the Box Size 
  \label{size}}
\tablewidth{0pt}
\tablehead{
\colhead{Model} & \colhead{size} &
\colhead{$\overline{\< E_B \>}/\rho_{0}c_s^2$} & \colhead{${\lambda_{B,{\rm maj}}} \over{H}$} \\
\colhead{} & \colhead{} & \colhead{} & \colhead{} }
\startdata
y64      & $(1, \pi, 1) H$     & 0.024  & 0.36  \\
y64.x2   & $(2, \pi, 1) H$     & 0.028  & 0.49  \\
y64.y2   & $(1, 2\pi, 1) H$    & 0.035  & 0.45  \\
y64.x2y2 & $(2, 2\pi, 1) H$    & 0.038  & 0.49  \\
y64.x4y4 & $(4, 4\pi, 1) H$    & 0.038  & 0.57   \\ 

\enddata
\end{deluxetable}

\begin{figure}
\plottwo{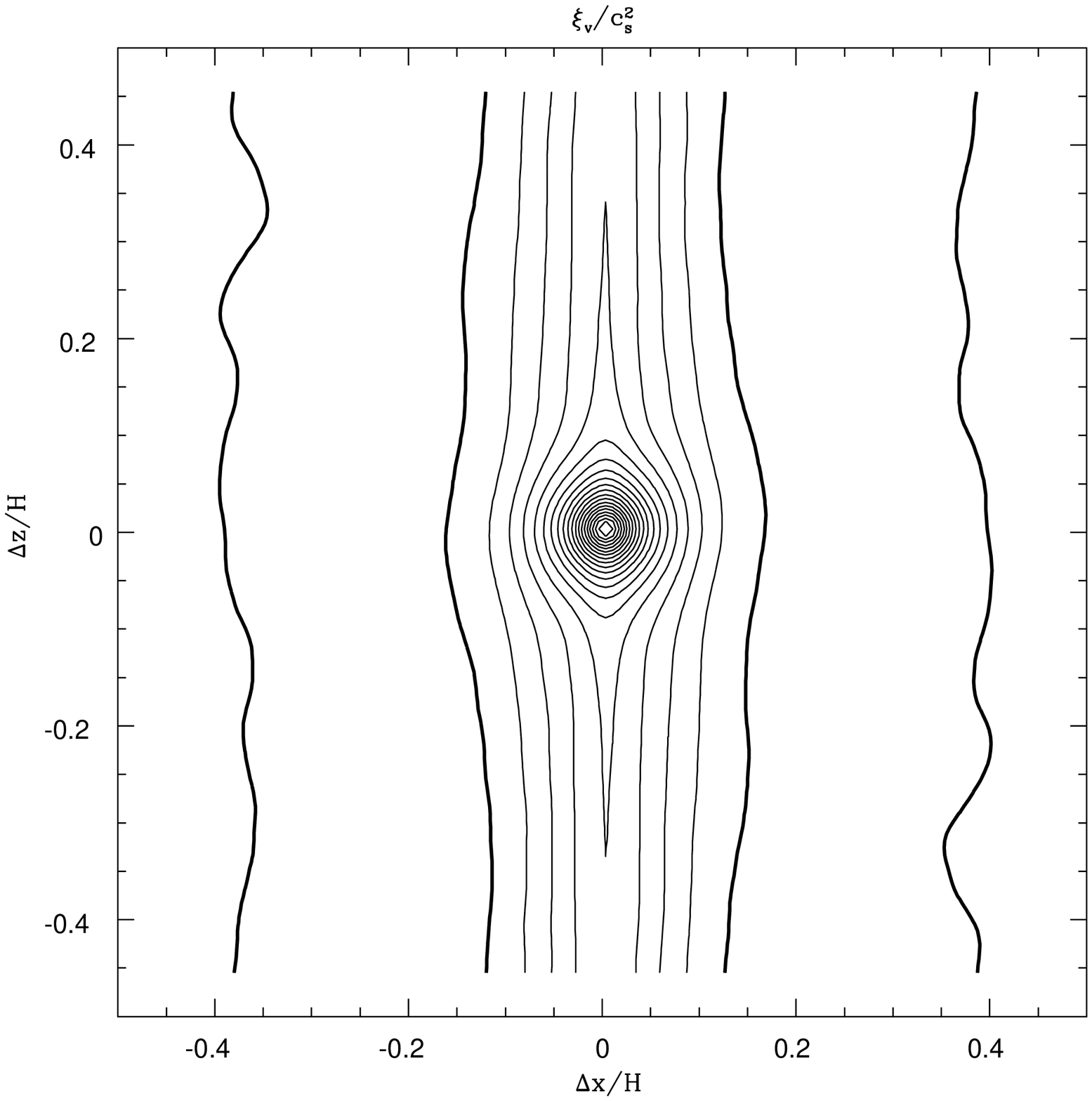}{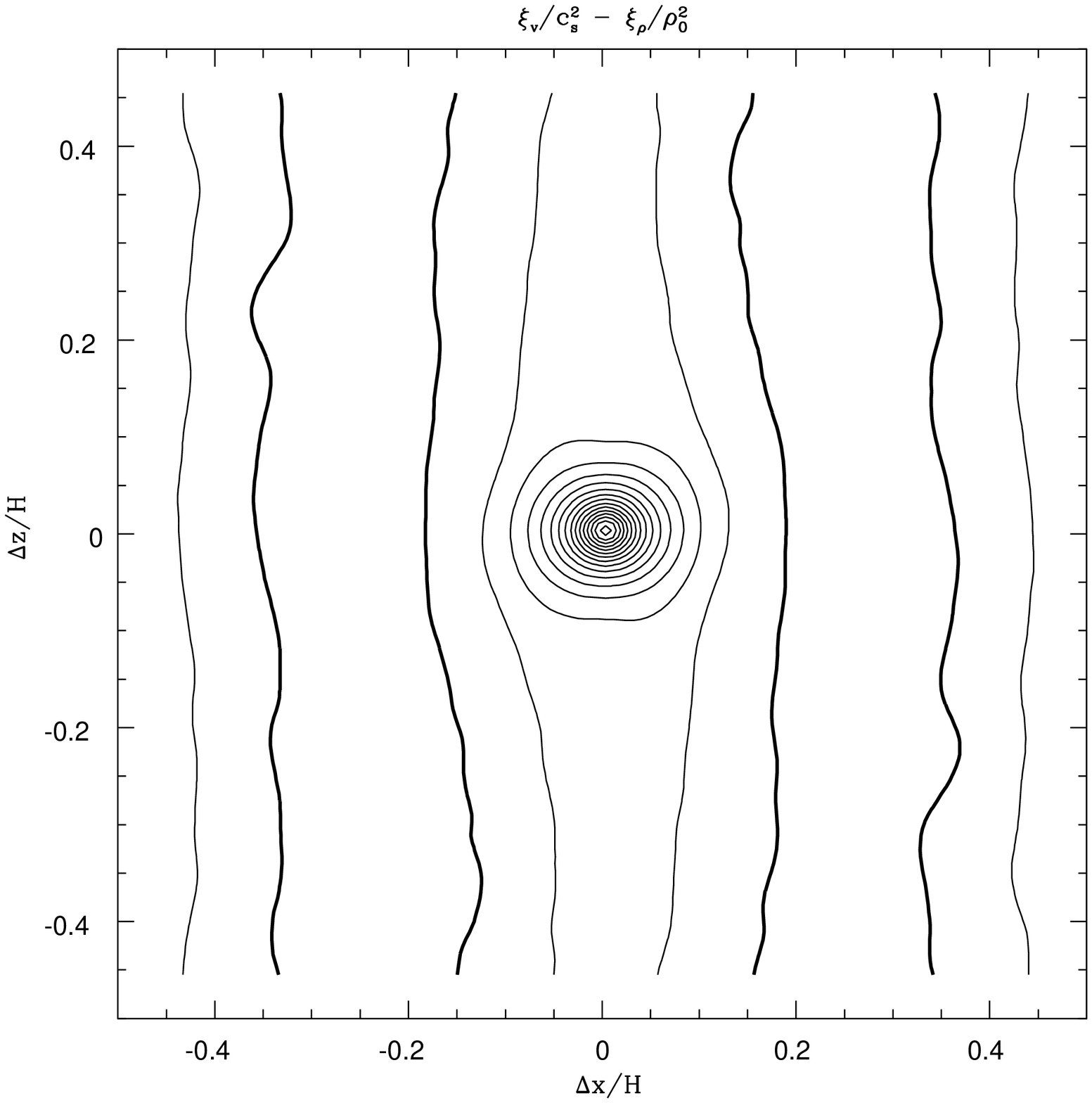}
\caption{
Turbulent velocity correlation function and a differential correlation
$\xi_v /c_s^2 - \xi_\rho/\rho_0^2$ in the $\Delta y = 0$ plane for
run $z128$.  In $\xi_v$, apart from the compact core at small
separations, there is a weak correlation at large scales that is likely
due to the sound waves. The contours run linearly from $0$ to
$0.005$ for $20$ levels. The heavy line is the $0$ contour.}
\label{fig:std.corr.xz}
\end{figure}

\begin{figure}
\plotone{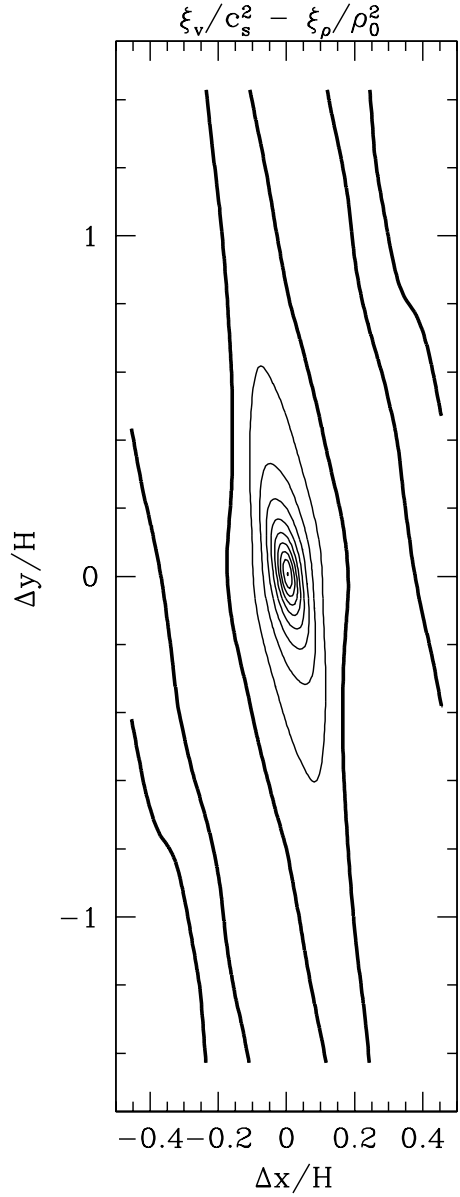}
\caption{
The differential correlation function $\xi_v/c_s^2 -
\xi_{\rho}/\rho_0^2$ in the $\Delta z = 0$ plane in run $z128$.
After removing the contribution due to the sound waves, the correlation
ellipsoid is more compact and almost identical to that of the magnetic
field. The contour levels are set the same as in Figure
\ref{fig:std.corr.3fig}, linearly from $0$ to $0.009$.}
\label{fig:std.soundwave}
\end{figure}

\begin{figure}
\plotone{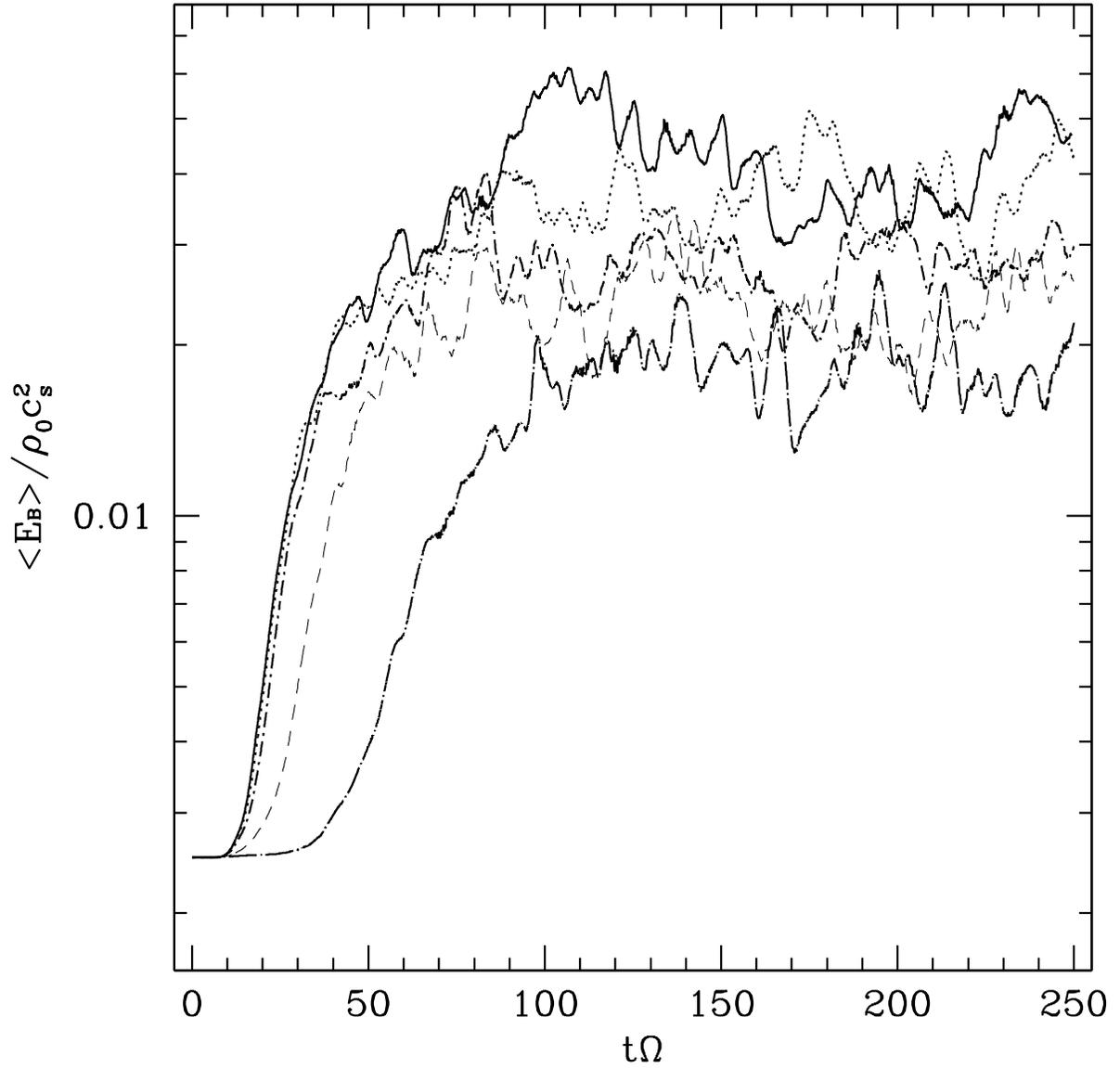}
\caption{
Evolution of the magnetic energy in the net azimuthal field run.  From
bottom to top the lines are: heavy dot - long dash: $32/H$; grey short
dash: $64/H$; heavy dot - short dash: $128/H$; grey dot: $192/H$; heavy
solid: $256/H$.  The saturation energy increases with resolution. 
}
\label{fig:mb.eb.vs.t}
\end{figure}

\begin{figure}
\epsscale{0.4}
\plotone{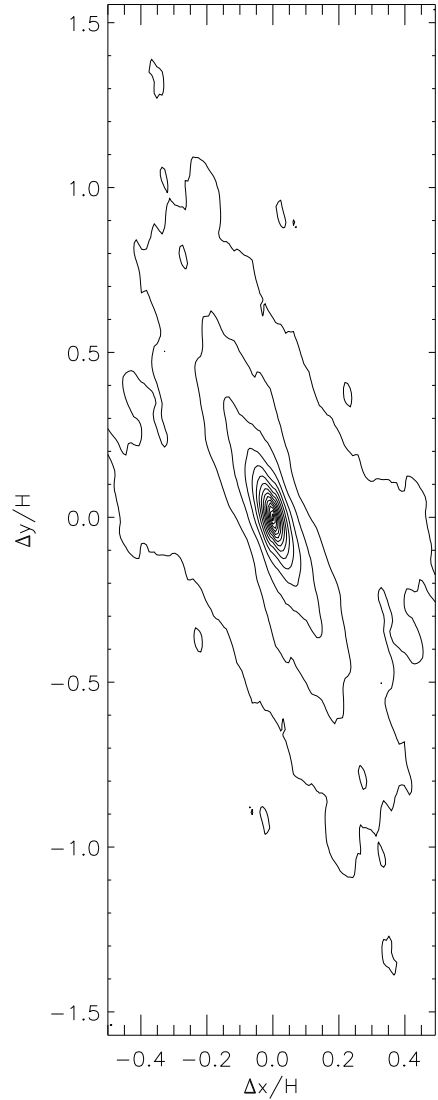}
\caption{
Correlation function in the $\Delta z = 0$ plane for the ``field line
stretching'' term in the magnetic energy equation.  The data is from the
$y128b$ \athena run and the contour levels run logarithmically from
$10^{-5.1}$ to $10^{-3.6}$.  The generation and dissipation of
magnetic energy occurs in a local manner, consistent with the
localization of the dynamical variables.  
}
\label{fig:me.xy.no_shear}
\end{figure}

\begin{figure}
\plotone{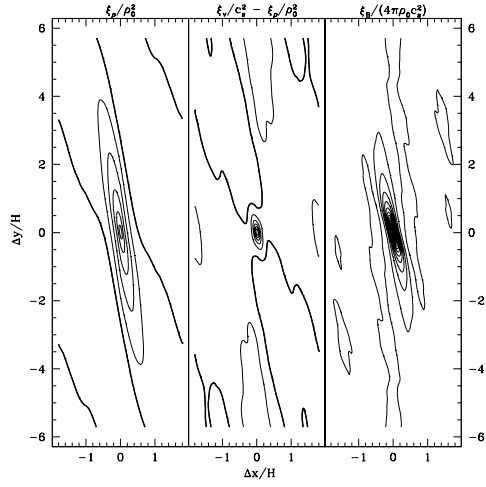}
\caption{
The density correlation function, a differential correlation
$\xi_v/c_s^2 - \xi_{\rho} /\rho_0^2$ and the magnetic correlation
function in the $\Delta z = 0$ plane for run $y64.x4y4$. The
contours are set linearly from $-0.007$ to $0.08$ for $20$ levels. The
heavy line is the $0$ contour.}
\label{fig:size.soundwave}
\end{figure}

\end{document}